# Coherent Bunching of Anyons and Dissociation in an Interference Experiment

Bikash Ghosh[1], Maria Labendik[1], Vladimir Umansky[1],

Moty Heiblum[1,#], and David F. Mross[2]

[1] Braun Center for Sub-Micron Research, Department ofCondensed Matter Physics, Weizmann Institute of Science,Rehovot 76100, Israel

[2] Department of Condensed Matter Physics, Weizmann Institute of Science, Rehovot 76100, Israel.

[#] Corresponding author

[#]Corresponding author. Email: **moty.heiblum@weizmann.ac.il**

**Aharonov-Bohm (AB) interference of fractional quasiparticles in the quantum Hall Effect generally reveals their elementary charge ($e^*$)[1-15]. Recently, our interferometry experiments with several 'particle states' reported flux periods of $\Delta\Phi=(e/e^*)\Phi_0$ (with $\Phi_0$ the flux quantum) at moderate temperatures[16]. Here, we report interference measurements of 'particle-hole conjugated' states at filling factor $\nu$=2/3, 3/5, and 4/7, revealing unexpected flux periodicities of $\Delta\Phi=\nu^{-1}\Phi_0$. The measured shot noise *Fano factor* (F) of the partitioned quasiparticles in each of the interferometer's quantum point contacts (QPCs), was found to be, F=$\nu$[17], and not that of the elementary charge, F=$e^*/e$[18,19]. These observations point to interference of *bunched* (clustered) elementary quasiparticles as coherent pairs, triples, and quadruplets, respectively. A small metallic gate ('top gate,' TG), deposited in the center of the interferometer bulk, forming an antidot (or a dot) when charged, thus introducing local quasiparticles at the (anti)dot's perimeter. Surprisingly, such charging led to a dissociation of the 'bunched quasiparticles' and thus recovered the conventional flux periodicity set by the elementary quasiparticles' charge. However, the shot noise Fano factor (of each QPC) consistently remained at F=$\nu$, possibly due to the neutral modes accompanied the conjugated states. The two observations – bunching and debunching (or dissociation) – are not expected by current theories. Similar effects may likely arise in Jain's 'particle states' (at lower temperatures) and at even denominator FQH states[20].**

Interference measurements of quantum particles offer detailed insights into single-particle dynamics and collective behavior. Such measurements are particularly valuable for emergent exotic particles that form due to strong interactions between electrons in two dimensions and under strong magnetic fields. These particles carry fractional charges, obey anyonic statistics, and produce specific macroscopic signatures such as (fractionally) quantized electronic and thermal Hall conductances.

The properties of individual fractional quasiparticles (QPs) exhibit striking deviations from those of weakly interacting electrons. Experiments were designed to measure the scattering phase shifts and quantum coherence times[21], test entanglement between pairs of quantum particles[22], and probe the quasiparticle charge[23-25]. The *braiding phase* was measured via interference experiments using Fabry–Perot interferometers[1-12] or Mach–Zehnder interferometers[13-15]. Recently, *time-braiding* was utilized to measure the anyonic statistical phase via cross-correlation of partitioned quasiparticles[26-29].

In any fractional quantum Hall (FQH) interferometer, the interference loop is supported by two QPCs (serving as 'optical beam splitters') and ballistic propagation along edge channels surrounding the bulk. The QPCs inject discrete charge quanta in a particle beam, carrying shot noise characterized by a *Fano factor,* F [23-25]. Only the fractional charges that exist as bulk quasiparticles can tunnel in the QPC opening, while the QPC transmission is high. The smallest fractional value is the elementary QP charge *e**, which is typically determined by the denominator of the bulk filling factor $\nu_b$, i.e., *e**=*e*/3 at fillings $\nu_b$ =1/3 and *2*/3, *e**=*e*/5 at $\nu_b$ =2/5 and *3*/5, and *e**=*e*/7 at $\nu_b$ =3/7 and *4*/7. Notably, the gapless edge or interface modes impose no constraints, and the interference period is independent of their choice [16,30]. If the tunneling of individual quasiparticles is responsible for the interference and noise, the Aharonov-Bohm (AB) flux periodicity is expected to follow the simple rule, $\Delta\Phi = \Phi_0/\mathrm{F}$.

Our measurements were performed with a chiral (optical-like) Mach-Zehnder Interferometer (OMZI)[16,31,32] (Fig. 1). Its flexible design, based on co-propagating '*interface modes,'* is free of the deleterious effects of Coulomb interactions and allow interfacing of different fractional fillings. In this interferometer architecture, two charged external gates define the interfering *interface* modes: one gate *depletes* carriers to filling $\nu_d$, while the other gate *accumulates* carriers to filling $\nu_a$. The edge conductances of the two interface modes are set by the fillings differences $\nu_b$-$\nu_d$ and $\nu_a$-$\nu_b$. Two QPCs, QPC1 and QPC2 in Fig. 1, partition the co-propagating interface modes to form a closed interference loop. The chirality of the interface modes prohibits back-reflection; consequently, all incoming charge must pass through the inteferometer, and the OMZI displays a pristine AB interference pattern, known as 'pajama'[16]. To obtain phase slips in the AB interference, quasiparticles are introduced in the bulk of the interferometer by charging a small top metallic gate (TG), forming an isolated *antidot* (or a *dot*) in the center of the interferometer (Fig. 1)[16]. Our experiments were conducted with two OMZI devices featuring slightly different QPC transmissions, with the fabricated internal area of both OMZIs being A=3μm$^2$ and a single path length of 3μm, while the top gates radia are 0.5 μm (OMZI1) and 0.35μm (OMZI2).

Previous OMZI studies of three particle-like bulk fillings $\nu_b$=1/3, 2/5, and 3/7, found AB interference with magnetic flux periodicities $\Delta\Phi \cong (e/e^*)\Phi_0$, set by the elementary quasiparticle charges at an electron temperature of 15-20mK[16]. Additionally, charging the top gate yielded phase slips with the theoretically expected magnitudes. Here, we present AB interference and noise measurements of *particle-hole conjugated* bulk fillings $\nu_b$=2/3, 3/5, and 4/7.

We initially kept the TG uncharged, $V_{TG}$=0, and measured the shot noise of a single QPC at all three bulk fillings and for different choices of $\nu_d$ and $\nu_a$ in OMZI1. We consistently obtained the Fano factor F=$\nu_b$ instead of the conventional F$\cong e^*/e$ (Fig. 2). These measurements agree with earlier studies in conventional QPCs, with both sides depleted, $\nu_d$=$\nu_a$, at a similar temperature range[17,33]. Next, we partially closed the second QPC to form the OMZI and measured AB interference. The observed magnetic flux periodicity was non-integer in all cases and followed the simple rule $\Delta\Phi = \Phi_0/\nu = \Phi_0/\mathrm{F}$ (Fig. 3). The most straightforward interpretation of noise and interference measurements is that bunches of two, three, or four elementary QPs tunnel coherently across the QPCs and interfere (see Supp. **S6** and the discussion below).

The expected periodicity in the 'modulation gate' voltage of the OMZI1, $V_{MG}$, can be derived as follows. The flux periodicity revealed by the modulation gate should also be $\Delta\Phi=\frac{1}{\nu_b}\Phi_0$. The gate repels the charge $\Delta Q=\mathbf{C}\times\Delta \boldsymbol{V}_{\mathbf{MG}}$ with the capacitance **C** assumed to be independent of $\nu_b$, leading to a filling-factor independent periodicity $\boldsymbol{\Delta V}_{\mathbf{MG}}\times\nu_b\times B=\boldsymbol{\Phi}_{\mathbf{0}}\times\boldsymbol{\Pi}$, with $\Pi=ne/\mathbf{C}$. We find nearly identical values of $\boldsymbol{\Delta V}_{\mathbf{MG}}$ (2/3, 3/5, 4/7)=(11.7mV, 10.3mV, and 9.5mV) in OMZI1.

Next, we charged the top gate of OMZI1 by $V_{TG}$=-100mV (forming an antidot) or by $V_{TG}$=+50mV (dot), anticipating phase slips. Unexpectedly, we found that the AB flux periodicity changed to $\Delta\Phi \cong (e/e^*)\Phi_0$ at all the three tested fillings, in a similar fashion to the observed AB periodicities of particle-like Jain states[16] [Fig. 4 (a, c, e)]. Even more surprisingly, the change in the AB periodicity was not reflected in the measured properties of a single QPC, whose transmission and Fano factor were not affected by the $V_{TG}$ (See Supp. **S6**). In particular, the rule $\Delta\Phi = \Phi_0/F$ was violated in this case. For some parameters, a significantly weaker FFT component of the *bunched periodicity* $\Delta\Phi \cong \nu_b^{-1}\Phi_0$ remained in addition [Fig. 4 (b, d, f)] (See Supp. **S9**). The OMZI2 was tested only at $\nu_b$=2/3, 3/5, and exhibited the same behavior (See Supp. **S11**).

The 'modulation gate' voltage periodicity in the regime of dissociated QPs also showed unexpected behavior. Assuming the same modulation gate capacitance and replacing $\nu_b$ by $e^*/e$, leads to $\Delta V_{MG} \times (e^*/e) \times B = \Phi_0 \times \Pi$, with $\Pi = ne/C$, the observed periodicities at $V_{TG} \cong$-100mV were $\Delta V_{MG}$(2/3, 3/5, 4/7)=(21.7mV, 17.4mV, 15.5mV). At $\nu_b$=2/3, the modulation gate periodicity doubled, just like the magnetic-field periodicity. However, the behavior at the other fillings differs significantly. While this behavior is not understood, we rule out the possibility that the Top Gate influences the capacitance: The residual $\Delta\Phi \cong \nu_b^{-1}\Phi_0$ peaks at $\nu_b$=3/5 remain at a constant modulation gate periodicity independent of $V_{TG}$ in both devices.

Finally, we show a systematic TG voltage dependence of the interference peaks at $\nu_b$=3/5, measured in OMZI2, and at $\nu_b$=2/3 in OMZI1, respectively. For a sequence of positive top gate voltages in the range, $V_{TG} \cong$0-70mV, at $\nu_b$=3/5, the flux periodicity changed continuously as the QPs dissociated *one-by-one* with the varying $V_{TG}$ (Fig. 5). It starts with the *bunched periodicity* $\Delta\Phi$=(5/3)$\Phi_0$ and ends at a *dissociated periodicity* $\Delta\Phi$=5$\Phi_0$ at $V_{TG}$=60m, while a weaker *bunched periodicity* remained. For $V_{TG}$>60mV, only the bunched periodicity $\Delta\Phi$=(5/3)$\Phi_0$ is visible once again.

Our observed bunching should be viewed in the context of earlier shot-noise measurements that found the Fano factor F=$\nu_b$ at low temperatures (~10mK) and F=$e^*/e$ at higher temperatures[20,33]. In standard QPCs with counter-propagating edges, the low-temperature noise Fano factor follows F=$\nu_b$. Moreover, the same Fano factor was measured on intermediate plateaus in the QPC transmission, where no charge partitioning of QPs occurs. Consequently, no noise is expected unless it is generated by neutral modes. A specific theoretical model based on stochastic equilibration at the edge was put forth in Ref. [30]

The present experiment focused on the low-temperature regime due to vanishing visibility above 60 mK (see Supp. **S10**). Intermediate transmission plateaus were not observed in the OMZI's QPCs with co-propagating edge modes; thus, the contribution of the neutral modes was not established. Moreover, the agreement between the Fano factor and AB periodicity through the relation $\Delta\Phi \cong \Phi_0$/F at $V_{TG}$=0 strongly suggests an incoherent origin of the increased Fano factor (due to neutral modes). Instead, the combined measurements suggest that coherent bunched QPs, formed by pairs, triplets, or quadruples of the elementary QPs, tunnel at the QPC. Specifically, the measurement reflects a partitioned and interfering QP charge of $e_{\Phi} = \nu e$. It is worth noting that $e_{\Phi}$ is also the amount of charge that accumulates when one additional flux quantum is threaded in the interferometer. The corresponding QP is sometimes referred to as a *Laughlin QP* (or as a *vortex*), distinct from the *elementary QP* with charge $e^*$ (the two coincide for Laughlin states such as $\nu = 1/3$).

Superficially, the observed bunching resembles the pairing and tripling in the integer quantum Hall regime[12,34,35]. Its origin is not well understood, despite theoretical efforts [36]. Several key aspects of the experiments distinguish these observations from the anyon bunching observed here, indicating that they have different origins. Most significantly, the pairing and tripling at integers were only observed when an inner edge mode formed a closed trajectory in the interferometer, and decoherence of the mode destroyed the effect. By contrast, no inner modes are present in the OMZI at zero top-gate voltage.

The most puzzling observation is the dissociation of the interfering QPs upon charging the Top Gate, indicated by the sudden change in the AB flux periodicity. In this case, the Fano factor of the shot noise in each of the QPCs disagrees with the AB periodicity, which suggests an incoherent contribution to the shot noise. A consistent physical picture is that depending on the TG, either Laughlin QPs or fundamental QPs are partitioned in the QPC. Tunneling of elementary QPs with charge $e^*$ activates neutral edge modes, which contribute incoherently to the Fano factor, leading to F=$\nu$, but do not change the interference. In contrast, tunneling of Laughlin QPs does not excite neutral modes at the standard edge with vacuum (see Supp. **S12**). Moreover, Ref. [30] argued that additional noise due to neutral modes does not change when edge modes are replaced by interface modes. Consequently, the Fano factor, and the AB periodicity are both set by the tunneling charge, $e_{\Phi} = \nu e$.

Our observations present a new paradigm: At low-temperature, the partitioned and interfering quasiparticles are bunched elementary quasiparticles. Moreover, introducing localized QPs by charging the top gate in the interferometer bulk leads to phase slips in particle-like states[16]; here, it dissociates the bunched elementary QPs. The mechanism behind these observations remains unclear. Localized QPs are expected to produce phase slips in the interference pattern (of bunched or dissociated elementary QPs), but none were observed at any top gate's voltage. A recent FPI experiment in bilayer graphene also reported an absence of phase slips in 'particle-hole conjugated' states, which was attributed to fluctuations in the number of bulk QPs [37]. It is possible that similar fluctuations in the number of QPs under the top gate also cause debunching, which is observed at the same TG voltage range where phase slips were observed in particle-like states[16].

Our findings highlight the important of studying of particle-like Jain states at very low temperatures where the Fano factor is governed by $e_\Phi$ instead of $e^*$ [20]. Firstly, we expect the flux periodicity to change at low temperatures. Secondly, tracking the previously observed phase slips in the low-temperature regime may provide additional vital insights into bunching, dissociation, and how they are influenced by adding localized QPs (at will) into the bulk of the interferometer. Control over the interfering quasiparticles will be critical in the search for non-Abelian statistics since the Laughlin quasiparticles are Abelian even in non-Abelian quantum Hall states such as the Moore-Read Pfaffian[38].

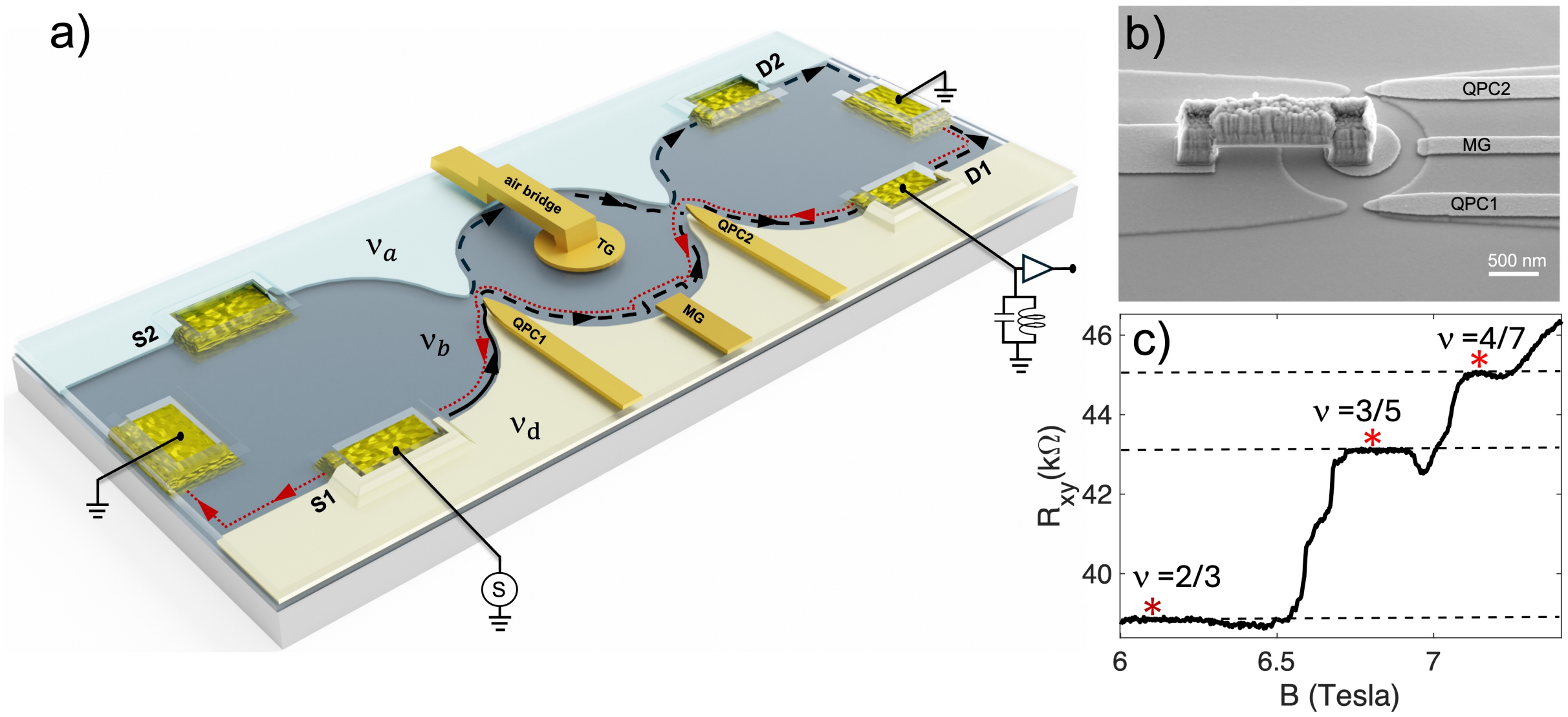

**Figure 1. Device structure of an optical Mach-Zehnder interferometer (OMZI).**

**(a)** Schematic representation of the OMZI. Its structure is based on three regions, each tuned to a different filling factor. The central region (the bulk, dark gray) is tuned by the magnetic field *B* to the filling $\nu_b$. It is flanked by two boundary regions, each tuned to different filling via the applied voltage to two wide metallic gates. The upper gate (light blue) is positively charged, thus *accumulating* charge, leading to a higher filling, $\nu_a > \nu_b$; the lower gate (light yellow) is negatively charged, thus *depleting* the charge, leading to a lower filling, $\nu_d < \nu_b$. This combination results in two co-propagating *interface charge modes* with 'effective fillings': $'\nu_a - \nu_b'$ and $'\nu_b - \nu_d'$, respectively. This design prevents backscattering of the interface modes and is free of the deleterious effects of the Coulomb interactions. Current is sourced from contact S1 (or S2) at a frequency of 766kHz. The current is partitioned at QPC1 into two forward-propagating chiral interface modes partitioned further by QPC2. The transmitted current is collected by drains D1 and D2, is filtered by an LC circuit (resonance at 766kHz), and is amplified by a Cold Amplifier. A small metallic 'top gate' (TG) is deposited at the center of the interferometer bulk, inducing an antidot (or a dot) that introduces localized quasiparticles at its perimeter. **(b)** A scanning electron micrograph image of the heart of the interferometer. **(c)** The transverse Hall resistance as a function of the magnetic field showing quantized plateaus at the fractional fillings under study. The star marks(*) on each plateau indicate the tuned magnetic field, *B*, where interference and shot noise are measured.

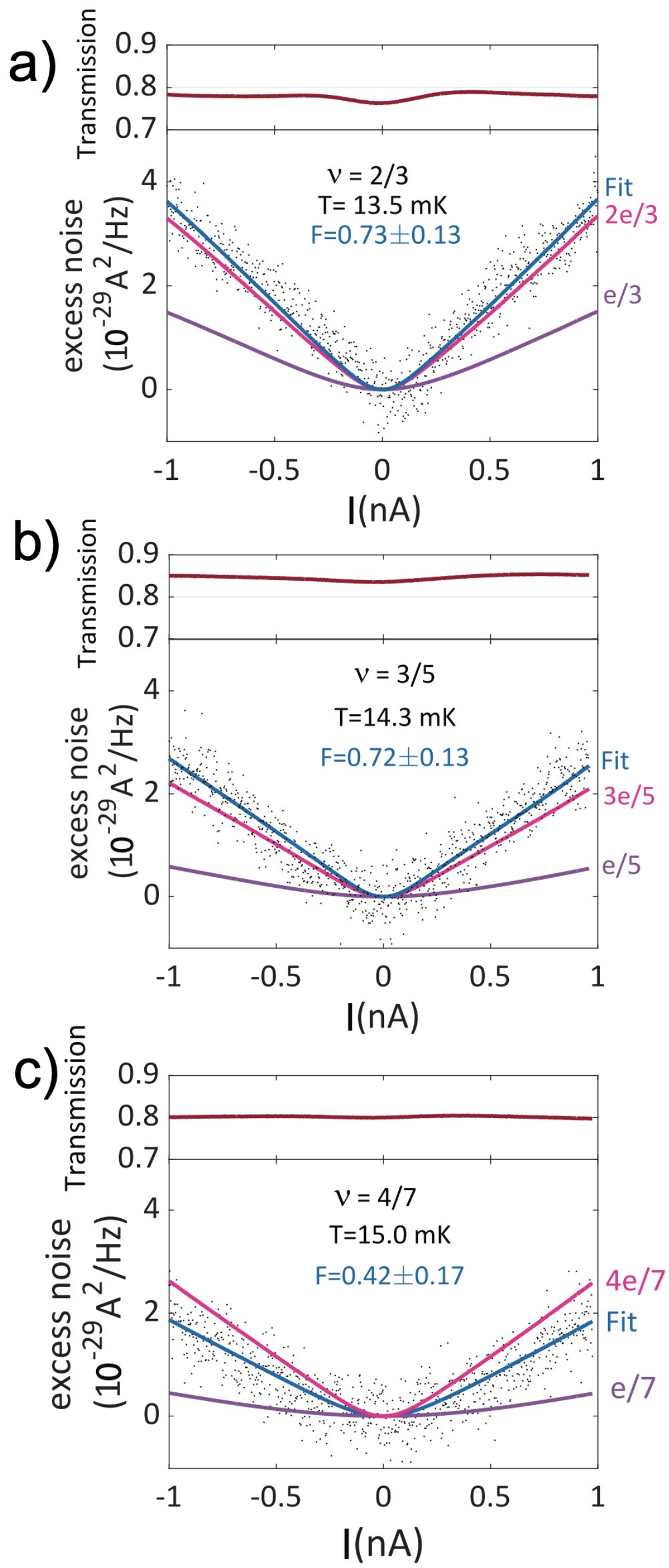

**Figure 2. Shot noise measured in a single QPC of the OMZI**. Noise measured at three 'particle-hole conjugated' states with bulk fillings: **(a)** $\nu$=2/3, **(b)** $\nu$=3/5, **(c)** $\nu$=4/7. **Top panels:** Nonlinear transmissions of the QPC. **Bottom panels:** Excess noise as a function of the sourced current. The blue lines represent the fitted curves and the Fano factor, F. The pink lines correspond to F=$\nu_b$, and the purple lines correspond to F=$e^*/e$.

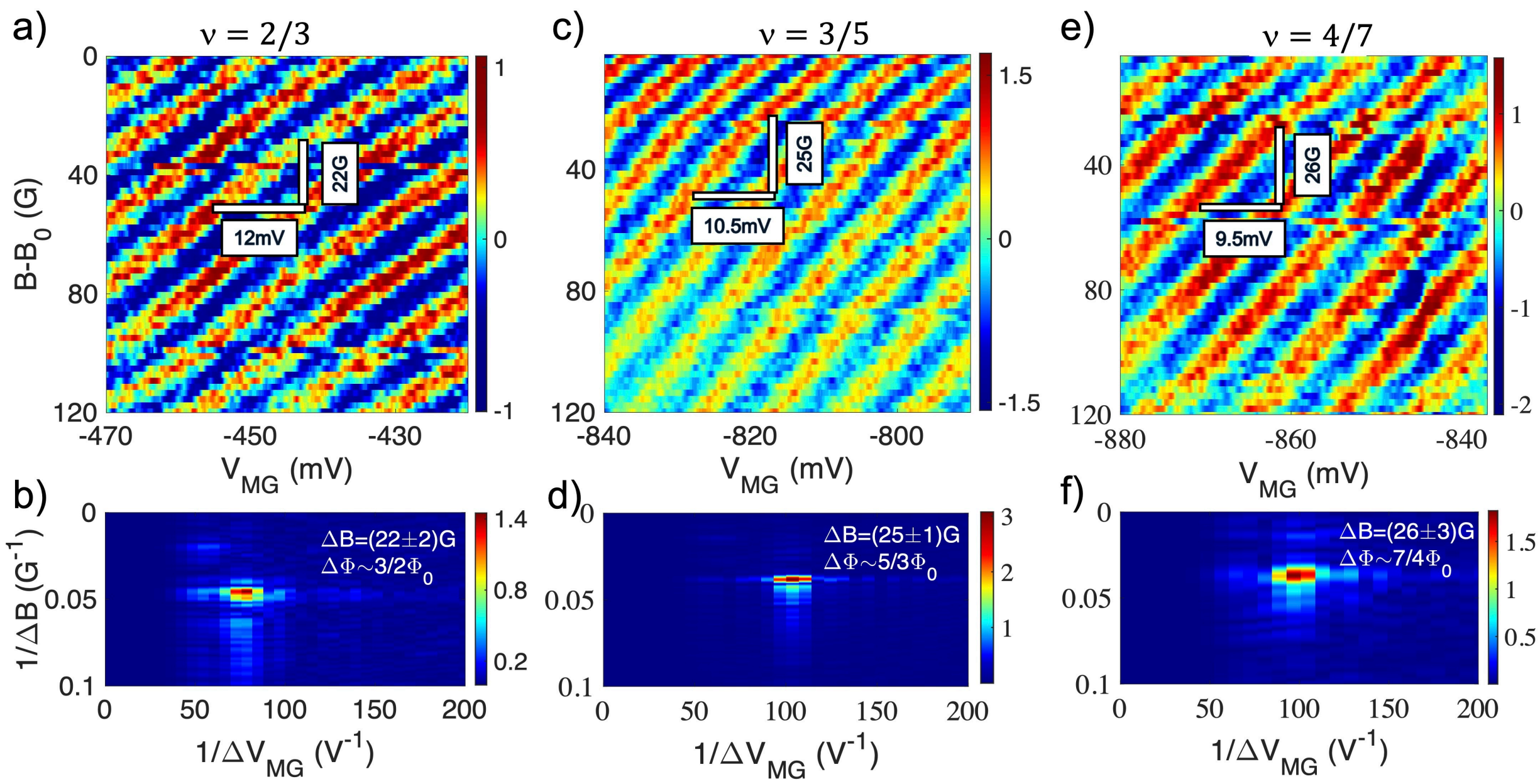

**Figure 3. Evidence of coherent bunching of anyons in particle-hole conjugated states.** Aharonov-Bohm (AB) interference (pajamas) of the measured three states, plotted as a function of magnetic field *B* and modulation gate voltage $V_{MG}$. Each of the two QPCs is tuned to transmission, *t*1=*t*2=0.8, while the top gate (TG) is not charged. The pajamas and the corresponding 2D Fast Fourier Transforms (FFTs) relate to the states: **(a, b)** $\nu_b$=2/3, **(c,d)** $\nu_b$=3/5, and **(e,f)** $\nu_b$=4/7. The fractional flux periodicities are $\Delta\Phi \cong \nu_b^{-1}\,\Phi_0$ in all three cases. The depleted and accumulated sides were at $\nu_d$=0 and $\nu_a$=1 for all bulk fillings.

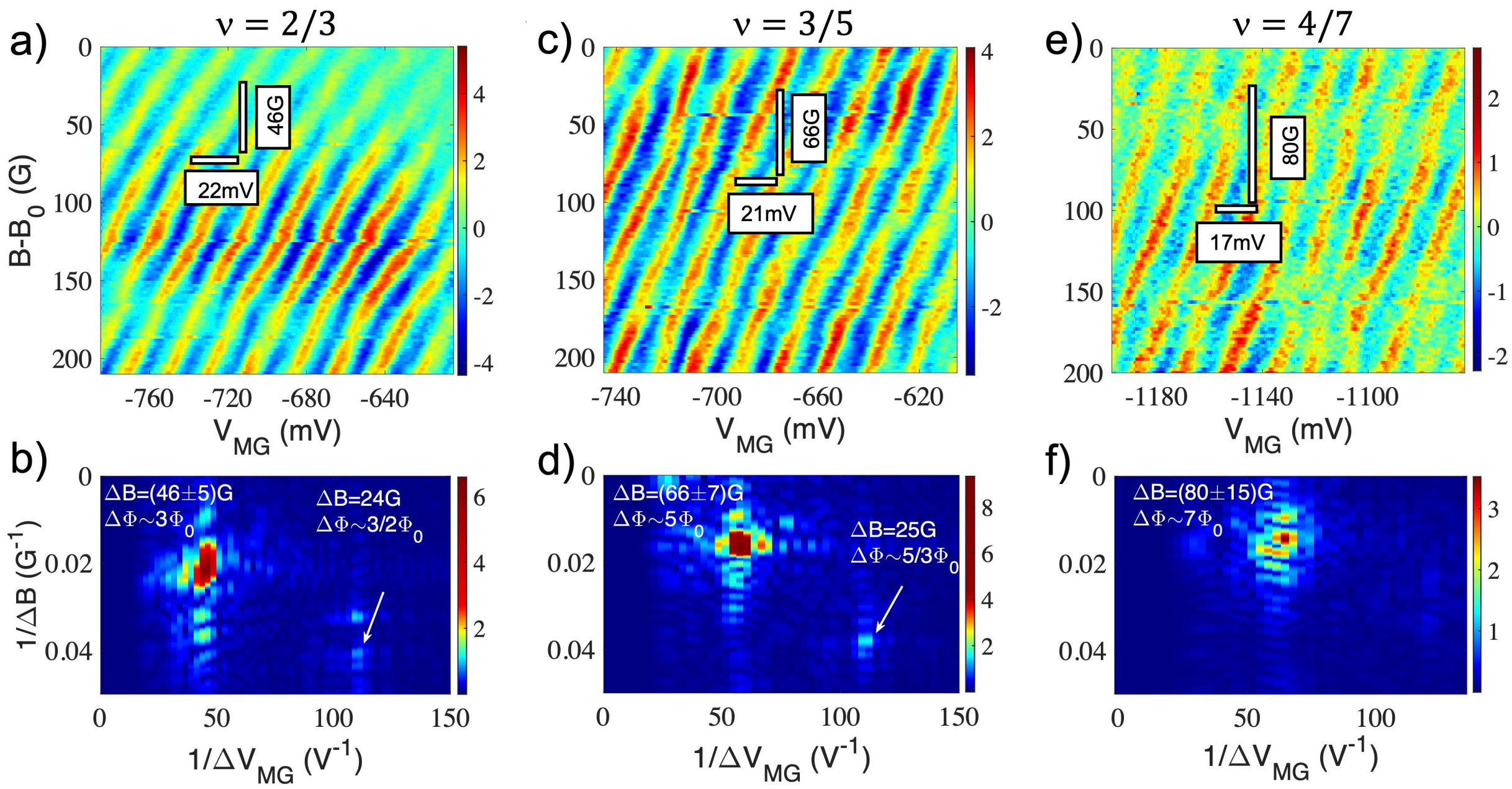

**Figure 4. Evidence of coherent dissociation of bunched anyons in 'particle-hole conjugated' states.** Activating the top gate with $V_{TG}$=-(100-120)mV forms an antidot that harbors isolated quasiparticles. An apparent *dissociation* of the bunched anyons takes place. Figures **(a), (c),** and **(e)** show typical AB pajamas in the $B$-$V_{MG}$ plane. The two QPCs' transmissions ($t1$=$t2$=0.8) are unaffected by $V_{TG}$. The FFTs in figures **(b)**, **(d)**, and **(f)** reflect the flux periodicities: $\Delta\Phi \cong (e/e^*)\Phi_0$, corresponding to the interference of the elementary fractional charges, $e/3$, $e/5$, and $e/7$. A weak higher harmonic due to residual bunching is also observable in $\nu_b$=2/3 and $\nu_b$=3/5; however, at filling $\nu_b$=4/7, the 'bunched periodicity' is absent.

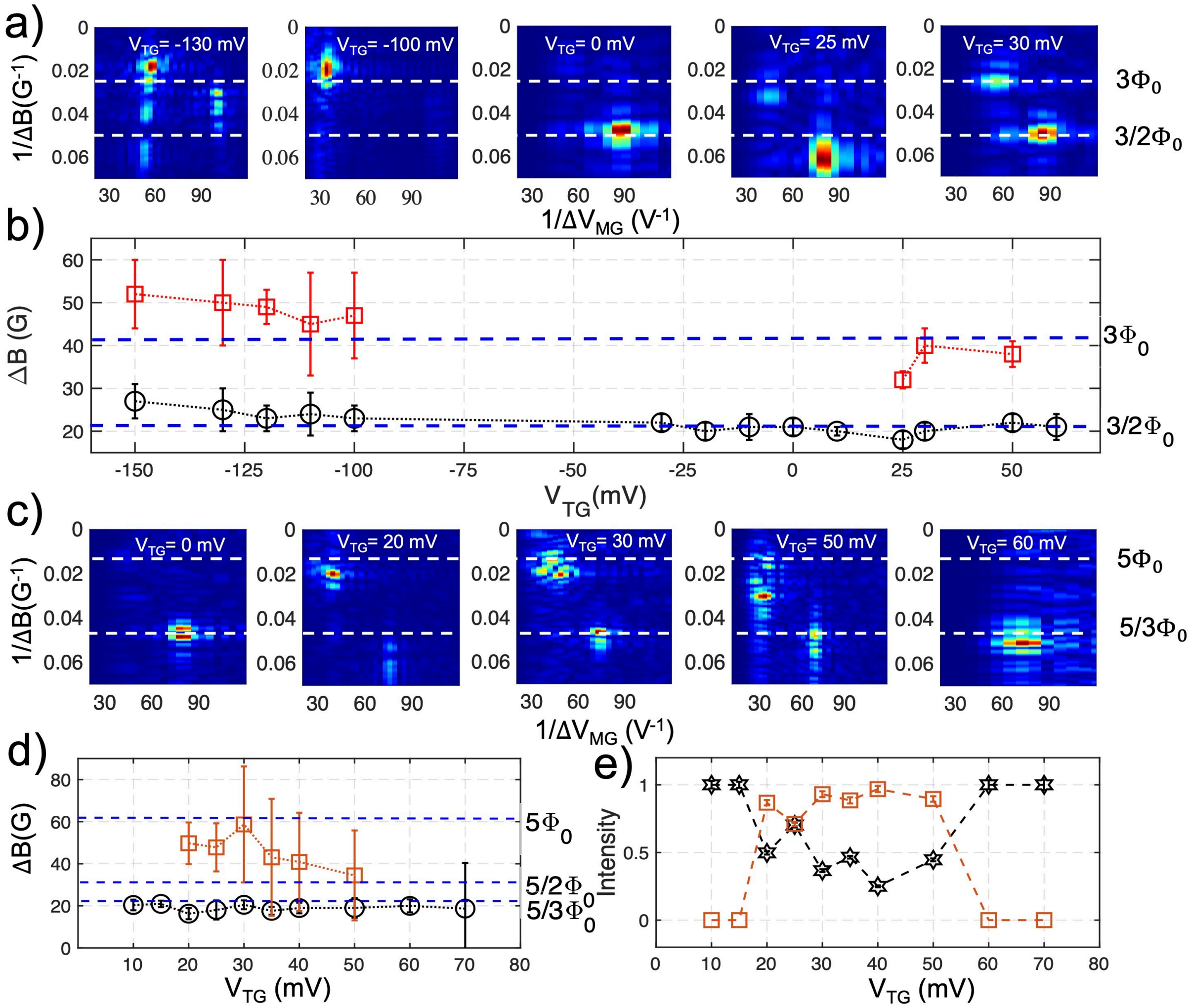

**Figure 5. Evolution of the dissociated anyons with charging the top gate at $\nu_b$=2/3 and 3/5**. The top gate (TG) was incrementally charged within different $V_{TG}$ ranges. For $\nu_b$=2/3, $V_{TG}$ was varied from -150 mV to +60 mV**,** while for $\nu_b$=3/5, the $V_{TG}$ incremented in steps of $\Delta V_{TG}$=5mV within the range (0-70)mV and the AB pajamas in the $B$-$V_{MG}$ plane were recorded at each step. Figures **(a)** and **(c)** show $\Delta B$ peaks in the 2D FFT as a function of $V_{TG}$ for $\nu_b$=2/3 and $\nu_b$=3/5. Anyon dissociation is observed with positive and negative TG bias (dot or antidot ). **(b)** QP dissociation (red) takes place in the range $V_{TG}$=-(150-100 )mV and $V_{TG}$=+(25-50 )mV for $\nu_b$=2/3, showing flux periodicity $\Delta\Phi=3\Phi_0$, while a residual weaker peak remains. QPs remain fully bunched between -30 mV and 25 mV with $\Delta\Phi=(3/2)\Phi_0$. We do not have any data from -30mV to -100mV. **(d)** For $\nu_b$=3/5, with $V_{TG}$ below 20mV or above 50mV, the QPs remain bunched, with flux periodicity $\Delta\Phi=(5/3)\Phi_0$. An apparent dissociation (red) of the bunched anyons (black) takes place in the range $V_{TG}$=(20-50)mV, showing partial bunching $\Delta\Phi=(5/2)\Phi_0$ and complete

dissociation $\Delta\Phi=5\Phi_0$ **(c,d).** A residual bunched peak remains. The error bars indicate the width of each peak, determined from a Gaussian fit. **(e)** Relative weights of the bunched (black) and dissociated (red) FFT peaks as a function of $V_{TG}$ for $\nu_b=3/5$.

## Methods

### Sample fabrication process.

The MESA of the device, with dimensions 250×650 $\mu m^2$, was prepared in a GaAs/AlGaAs heterostructure (see **S1**) by wet etching in $H_2O_2$: $H_3PO_4$: $H_2O$ =1:1:50 for 100 seconds. The 2DEG was located 170nm below the surface. Ohmic contacts were deposited at the edge of the MESA and within the MESA area (for testing the interface modes). The sequence of the evaporation by order of deposition is: Ni (15 nm), Au (260 nm), Ge (130 nm), Ni (87.5 nm), Au (15 nm). Contacts were annealed at 440℃ for 80 seconds. The sample was covered by 30nm of an insulating $HfO_2$, followed by evaporation of large metallic gates, Ti (5 nm) and Au (15 nm). The charged gates 'separate the device area' into three regions with fillings: $\nu_a$, $\nu_b$, $\nu_d$. The MESA was covered by 25 nm of $HfO_2$ layer, followed by a deposition of quantum point contacts (QPCs), a modulation gate (MG), and a small top gate (TG) in the middle of the bulk, with a radius of 0.5 µm (or 0.35 µm). In the final step, ohmic contacts, large metallic gates, the QPCs, and the modulation gate (MG) were connected to large pads via thick gold lines. The TG was connected by an *air bridge* and gold lines that passed over the $HfO_2$-coated metallic gates.

**Data availability** The data that support the plots within this paper and other findings of this study are publicly available at https://doi.org/10.5281/zenodo.15395284.

**Acknowledgments**
M.H. thanks Yuval Ronen for the fruitful discussions.
D.F.M. acknowledges many illuminating conversations on quantum Hall interferometry with Yuval Ronen.
B.G thanks Arup Kumar Paul for his helpful comment to improve device.
M.L. thanks the Ariane de Rothschild Women Doctoral Program for their support
D.F.M. acknowledges the support of the Israel Science Foundation (ISF) under Grant No. 2572/21 and by the Deutsche Forschungsgemeinschaft (DFG) within the CRC network TR 183 (project Grant No. 277101999).
M.H. acknowledges the support of the European Research Council under the European Union's Horizon 2020 research and innovation program. (Grant Agreement No. 833078) and the support of the Israel Science Foundation (ISF) (Grant No. 1510/22).

**Author contributions**
B.G. fabricated the devices. B.G. and M.L. performed the measurements and analyzed the data with the input from M.H. M.H. supervised the experiment's design, execution and data analysis. D.F.M. worked on the theoretical aspect and data analysis. V.U. grew the GaAs heterostructures. All authors contributed to the writing of the manuscript.

**Competing interests** The authors declare no competing interests.

## Supplementary Material

### S1. 2DEG structure design.

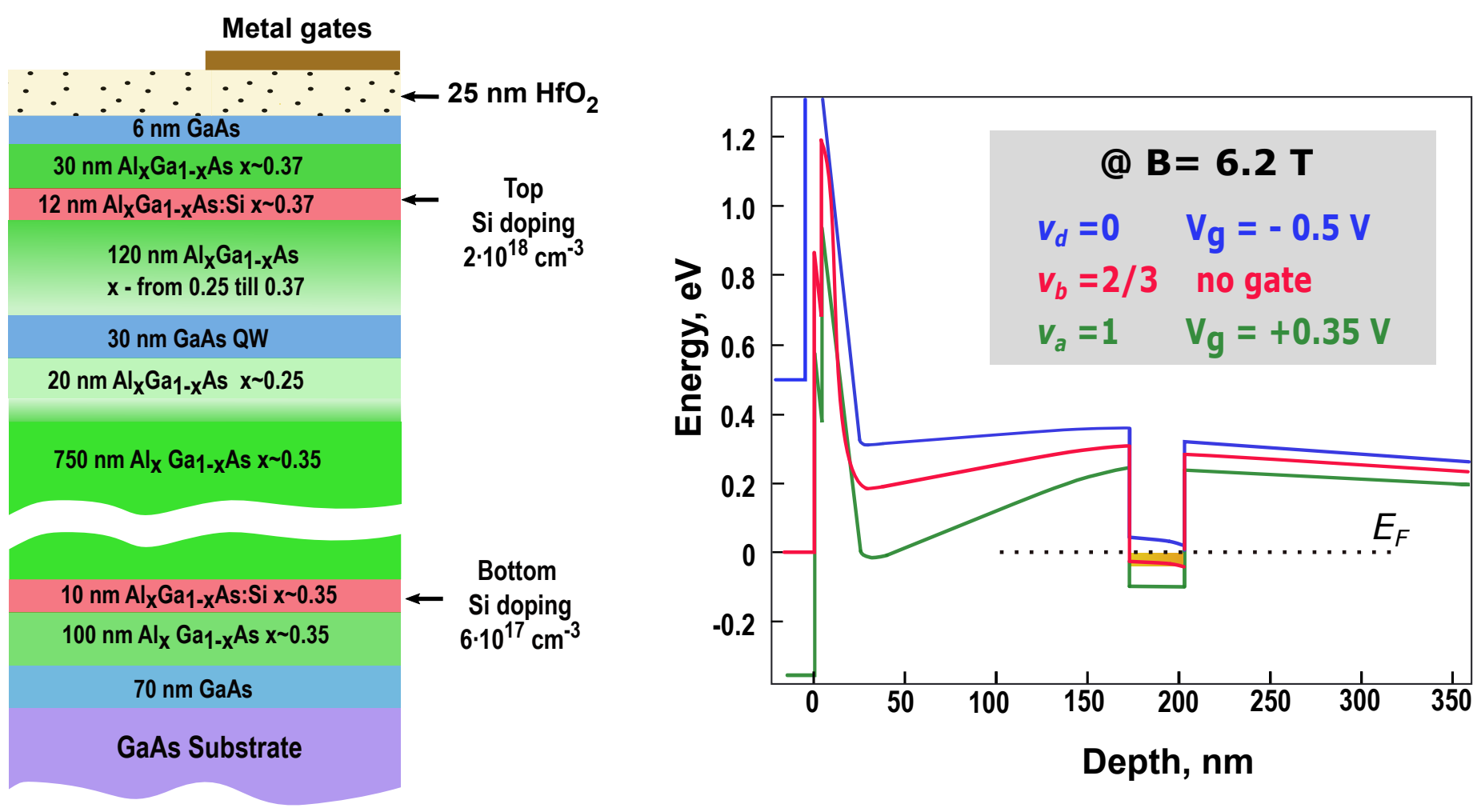

**Figure S1. The 2DEG within the heterostructure grown by MBE.** (Left) Cross section of the MBE growth (not in scale). (Right) The conduction band is depicted at two different gate voltages corresponding to two different filling factors $\nu_a$ and $\nu_d$. The filling factor $\nu_b$ is set by the magnetic field. The electron density at 4K is $1.07\times 10^{11}/\mathrm{cm}^2$ .

### S2. Interfacing 'particle-hole conjugated' states.

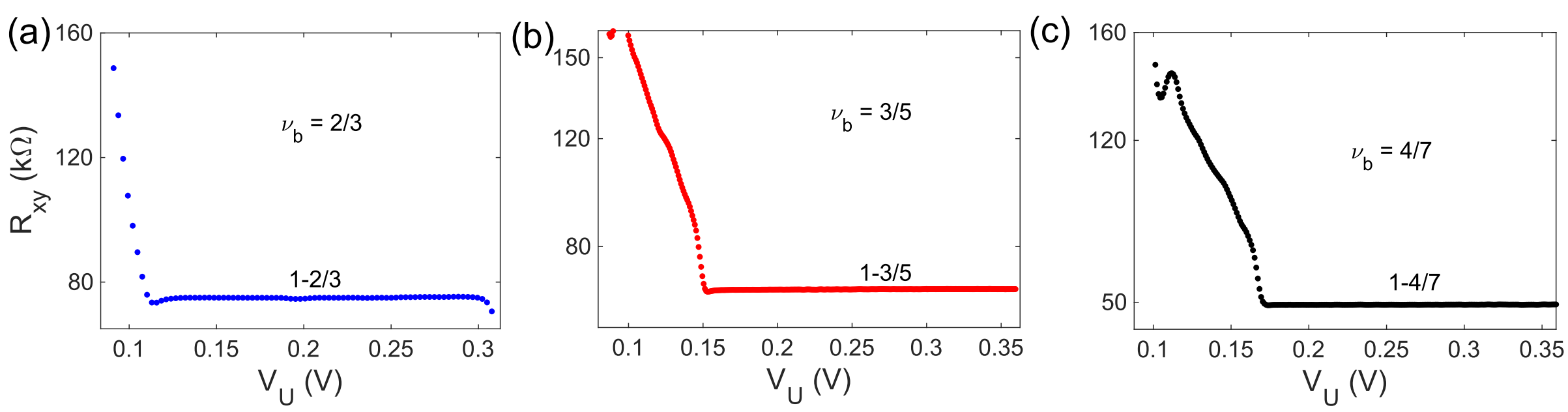

**Figure S2. Interfacing hole-conjugate states.** Two-terminal resistance measurements of the interface between $\nu_b$ and $\nu_a$. The bulk filling factor is fixed at $\nu_b$= 2/3 **(a)**, 3/5 **(b)**, and 4/7 **(c)**, while the accumulation gate is swept on the conductance plateau of $\nu_a$=1, with a positive bias. At $\nu_b$=2/3 and $\nu_a$=1, the 2DEG density underneath the gate increases by ~50%. Once charge equilibration is achieved, all interface modes move in the downstream direction.

**S3. Transmission of QPC in ‘particle-hole conjugated’ state at a ‘conventional QPC’.**

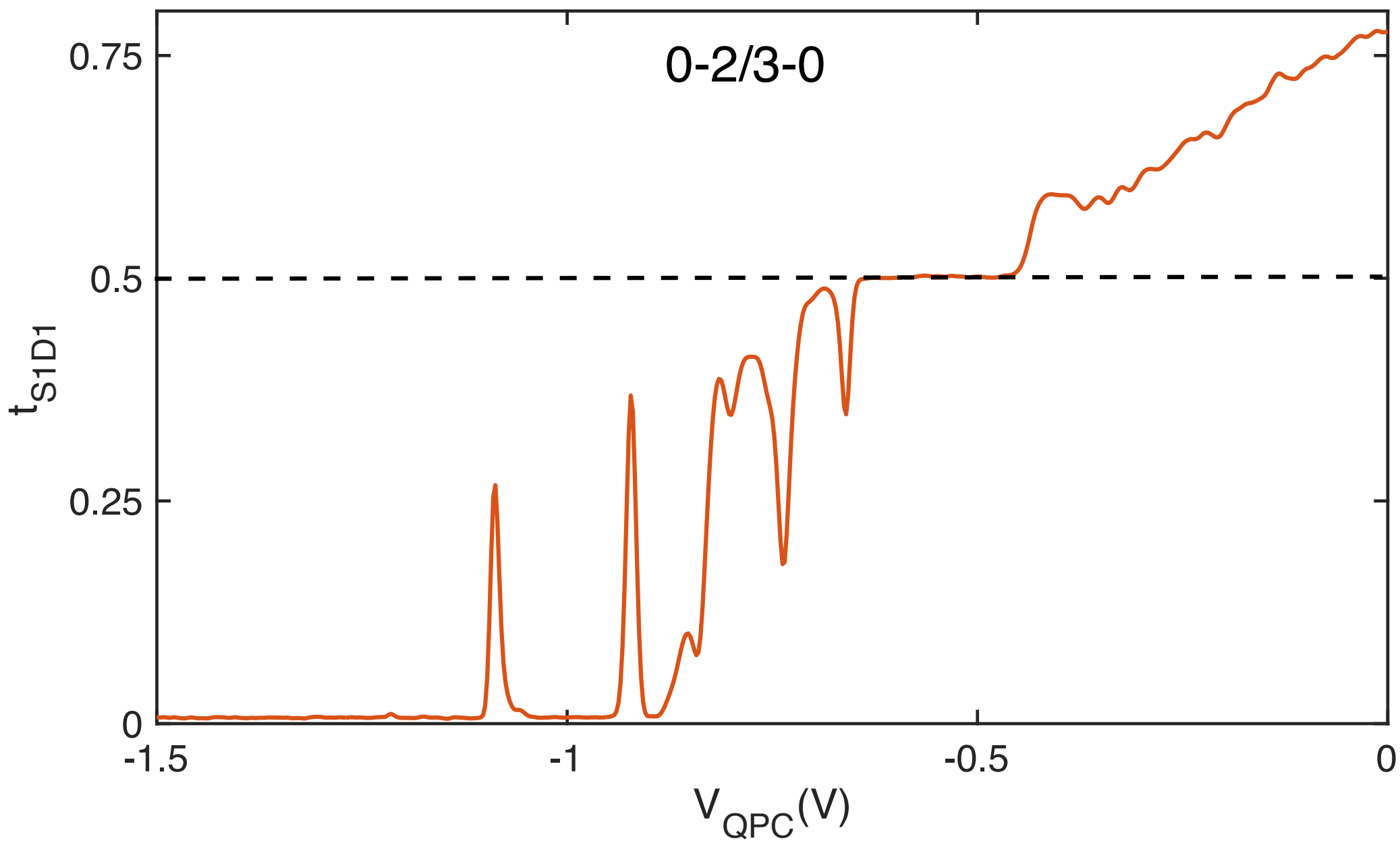

**Figure S3. QPC transmission in a ‘particle-hole conjugated’ state in the 0-2/3-0 configuration.** A constant transmission of t=1/2 (dashed black line) implies the formation of a $\nu_{QPC} = \frac{1}{3}$ conductance plateau in the QPC, and charge is not partitioned.

**S4. Transmissions of QPCs at OMZI of 'particle-hole conjugated' states.** In the OMZI setup, current can be sourced either from S1 or S2 and collected at drains D1 and D2 (See Fig.1). When the current ($I_{S1}$) is injected from S1, it is partitioned into two components, both of which are transmitted: i) $I_{S1D1}$, collected at drain D1 at the $\nu_b - \nu_d$ interface, and ii) $I_{S1D2}$, collected at drain D2 at the $\nu_a - \nu_b$ interface. The two transmission values are defined as $t_{S1D1} = I_{S1D1}/I_{S1}$ and $t_{S1D2} = I_{S1D2}/I_{S1}$.

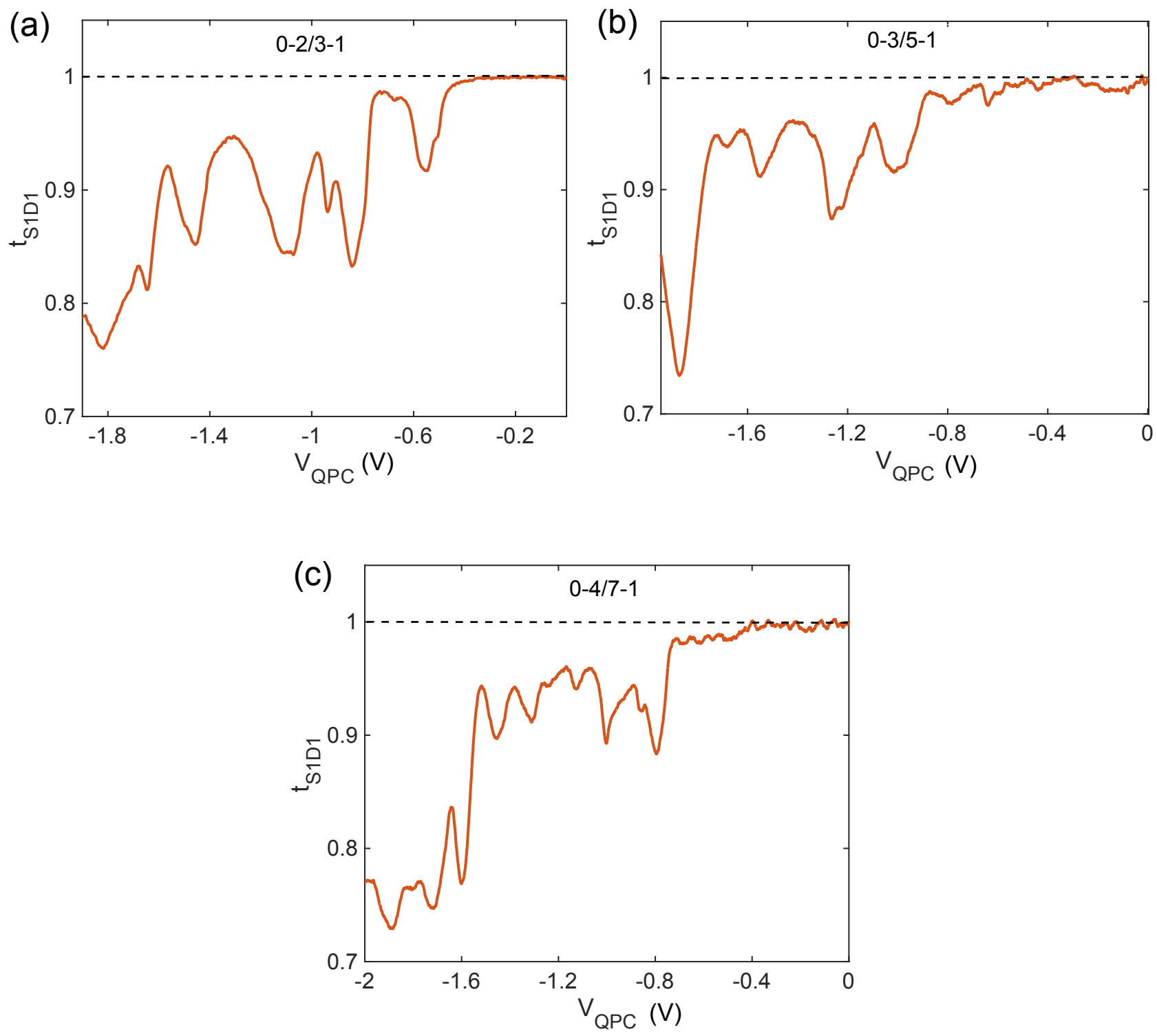

**Figure S4. Transmission at OMZI QPCs in particle-hole conjugated states.** One arm of the QPC is negatively biased, while the other arm is positively biased. Due to the chirality, the incident current must pass through the QPC – very different from a conventional QPC. We show the measured QPC transmission as a function of the QPC gate voltage for filling factor configurations: **(a)** 0-2/3-1, **(b)** 0-3/5-1, and **(c)** 0-4/7-1.

**S5. Calibration of the electron temperature.**

The fridge temperature was estimated using a calibrated $RuO_2$ resistor, which measured the temperature of the mixing chamber (MC) being approximately 12-15 mK. Since the electron temperature and fridge temperature may differ, a more accurate temperature measurement can be obtained through shot noise measurements at the known plateau of $\nu_{\mathrm{QPC}} = 1/3$ at $\nu_{\mathrm{bulk}} = 2/3$ in a conventional QPC (see Fig. **S3**).

The excess noise of a partitioned current at a finite temperature is well described by the analytic expression

$$S_I = 2FeIt(1-t)\left[\coth\left(\frac{FeIR_Q}{2kT}\right) - \frac{2kT}{FeIR_Q}\right].$$

Here, *I* is the impinging current, *t* is the transmission (ratio between the transmitted and impinging currents) of the QPC, $R_Q = \frac{1}{\nu_b}\frac{h}{e^2}$, is the resistance of the edge mode, *T* is the electron temperature, and *F* is the Fano factor. The fitting parameters are the temperature *T* and the Fano factor. Using the known value of Fano factor *F*=2/3 at the plateau of $\nu_{\rm QPC} = 1/3$ in $\nu_{bulk} = 2/3$ [26], the electron temperature extracted from the fit is T=13$\pm$2 mK.

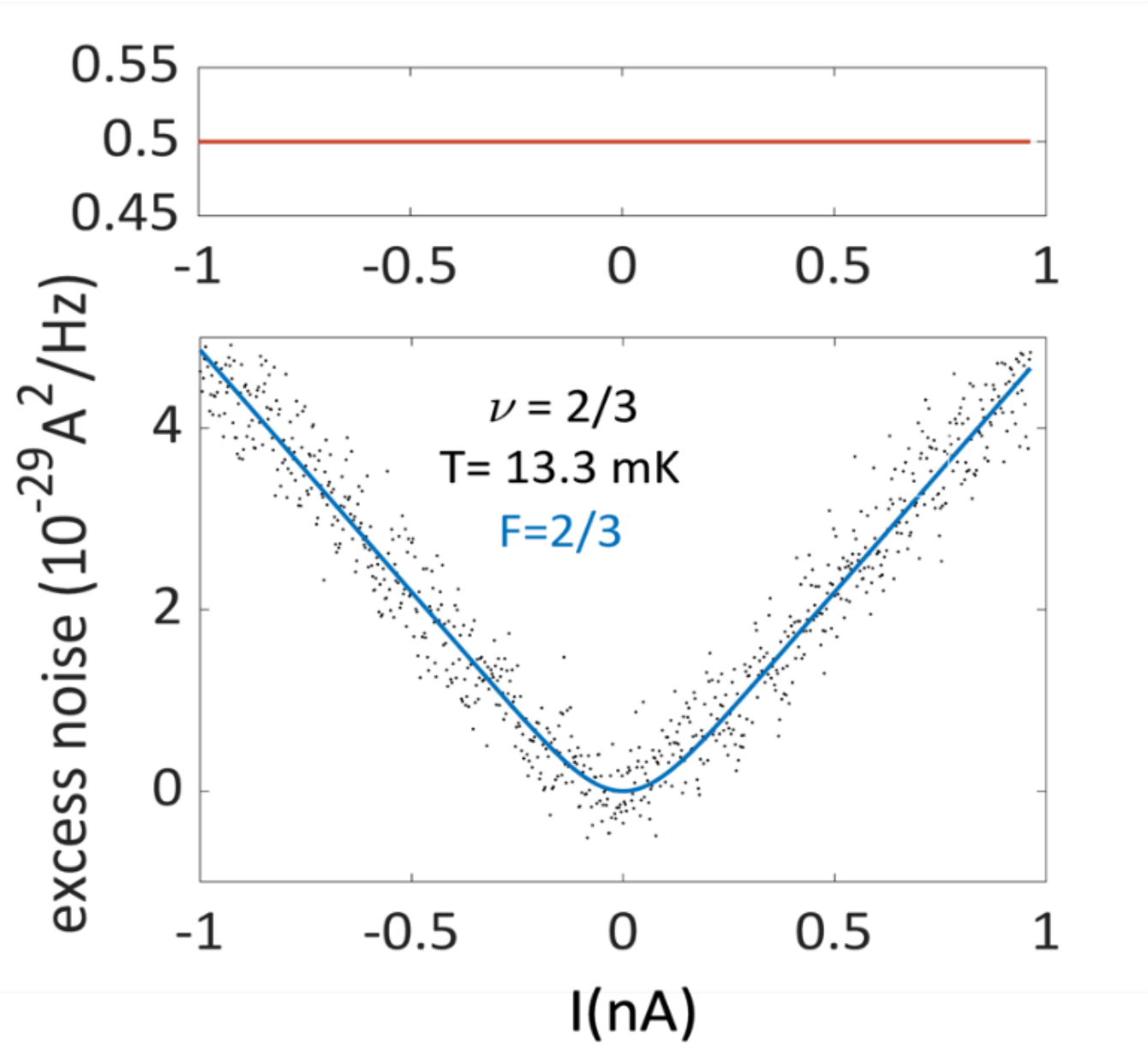

**Figure S5. Shot noise at $\nu_b = 2/3$ in 0-2/3-0 configuration.** Upper panel: QPC transmission at the $\nu_{\rm QPC} = 1/3$ plateau in $\nu_{\rm bulk} = 2/3$ (zero bias *t*=1/2) as a function of current. Lower panel: Excess noise as a function of bias current. Black dots are measured data, and the blue line is the fitted curve.

**S6. Effect of top gate on the QPC transmission and interferometer noise:**

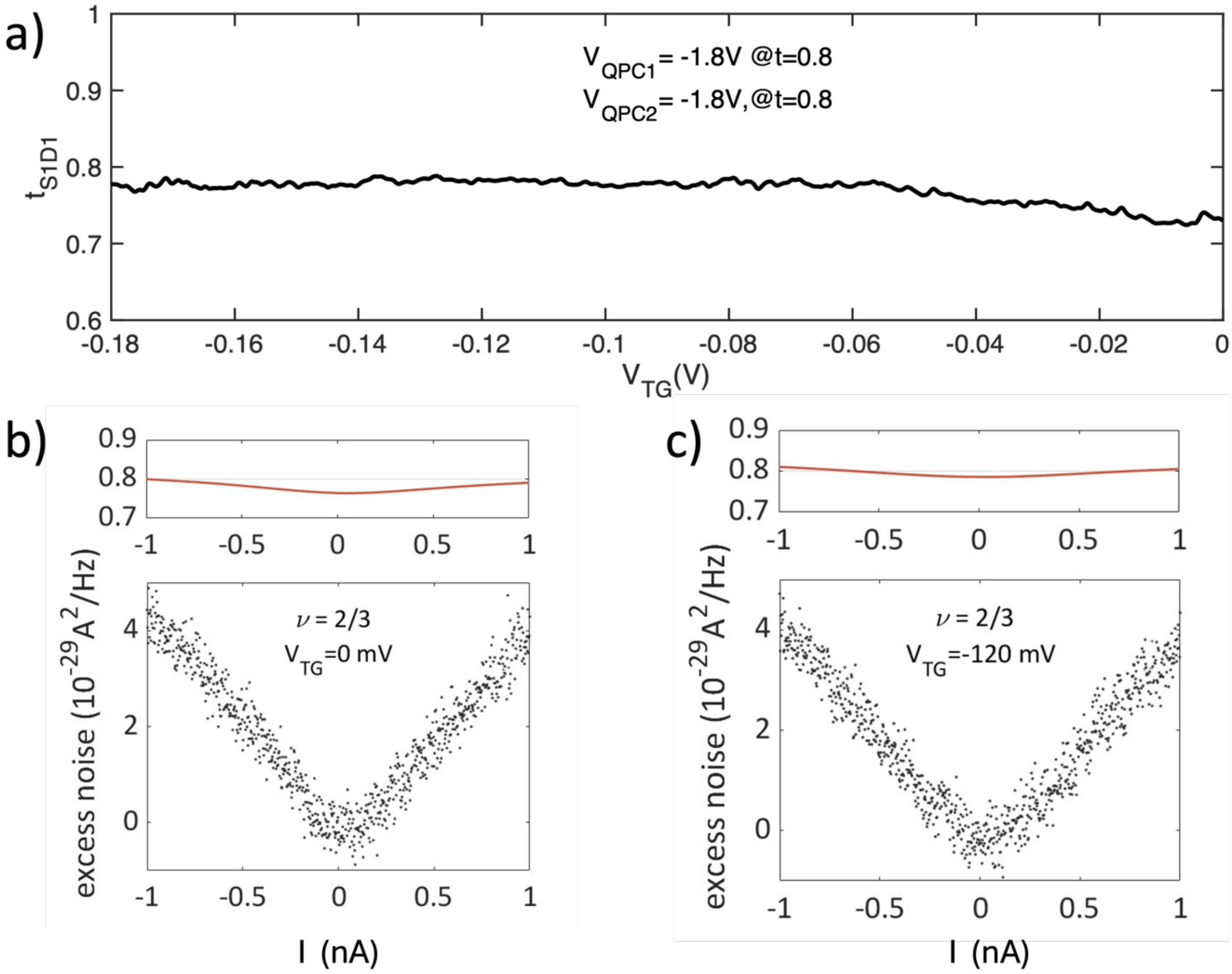

**Figure S6. Effect of the top gate on the QPC transmission and interferometer noise. (a)** Transmission of the interferometer as a function of the top-gate (TG) bias, with both QPCs set to a transmission of $t$=0.8. The shot noise of the entire interferometer at **(b)** $V_{TG}$ = 0 V and (c) $V_{TG}$ = -120 mV (within the dissociated regime) in the configuration 0-2/3-1. Neither the interferometer transmission nor the noise show a significant dependence on the TG bias. It is worth mentioning that the noise of the entire interferometer does not follow a simple relationship to the QP charge.

## S7. Dependence of the flux periodicity on the interface modes.

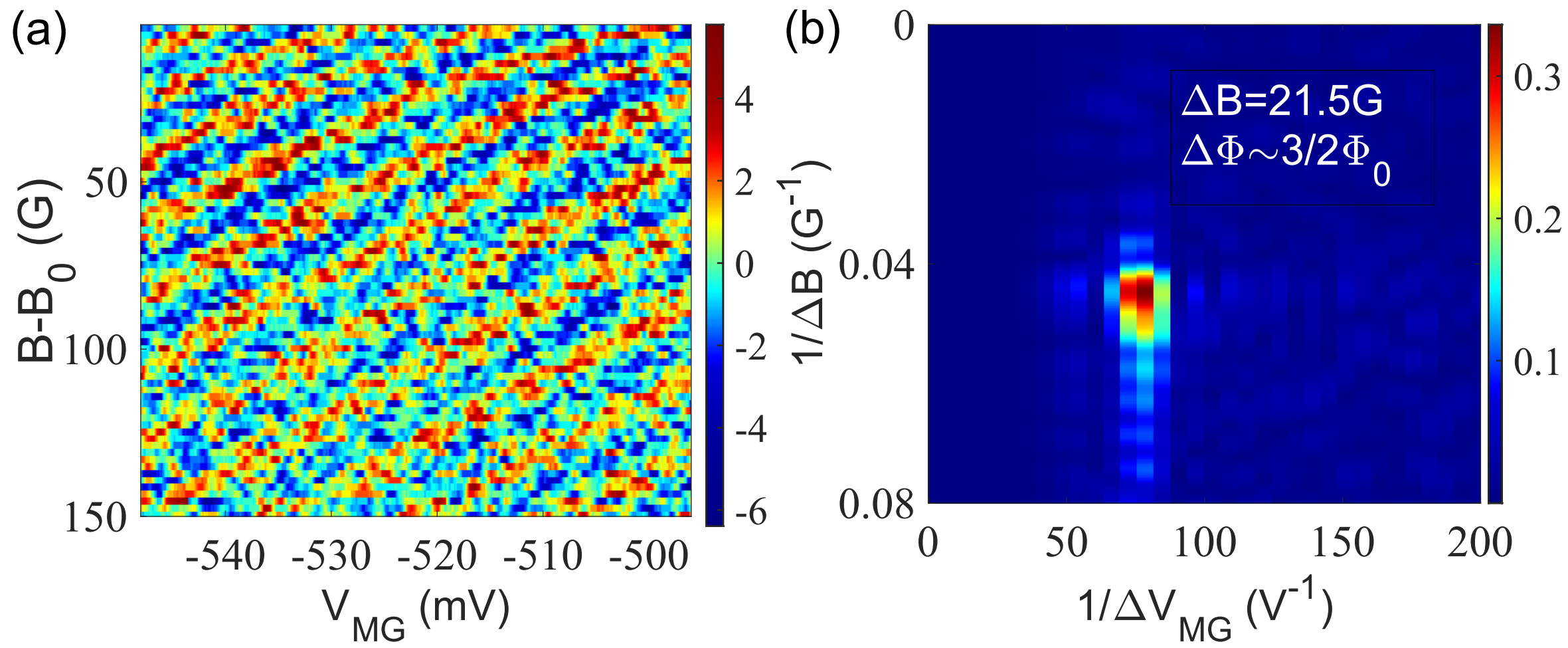

**Figure S7. Aharonov-Bohm interference configuration $\nu_a$=4/3, $\nu_b$=2/3, $\nu_d$=0. (a)** Conductance oscillations at $\nu_b$=2/3 in the $B$-$V_{\mathrm{MG}}$ plane, with 2DEG density underneath the accumulation gate increased by ~100%. The top gate (TG) in the bulk is not charged ($V_{\mathrm{TG}}$=0V). **(b)** The 2D FFT of the pajama plot shows a single peak with the periodicity $\Delta B$=21.5G, $\Delta V_{\mathrm{MG}}$=12.7mV. The flux periodicity is $\Delta\Phi \approx (3/2\}\Phi_0$, for an AB area of ~3μm². Similar results were observed in the fillings combination '1-2/3-0' (see main text) and '1-2/3-1/3', suggesting that the bulk filling solely determines the flux periodicity. The upstream neutral modes are different in the configuration 4/3-2/3-0 (downstream neutral at '4/3-2/3' and upstream neutral at '2/3-0') and in 1-2/3-0 (one upstream neutral mode at the '2/3-0' interface) and 1-2/3-1/3 (two upstream neutral at '2/3-1/3' interface), all suggesting that the fractional flux periodicity is not affected by the nature of the edge modes.

**S8. The effect of positive top gate potential on the bunched anyons**

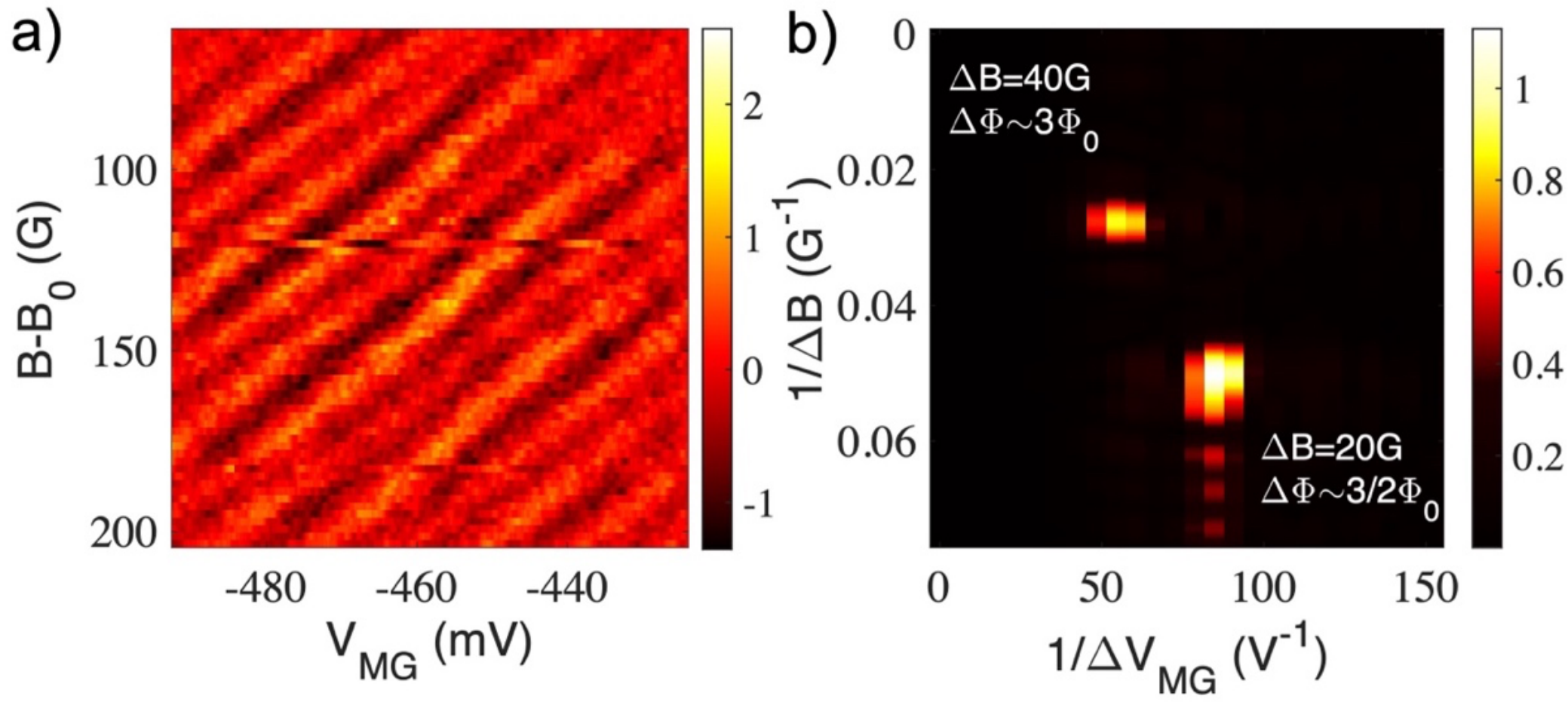

**Figure S8. Partial dissociation of bunched anyons via charging the TG: (a)** AB pajama with $\nu_b = 2/3$, and filling factors configuration: '1-2/3-0'. The QPCs are biased for transmission of 80%, and the top gate voltage is $V_{TG}$=+0.03V. **(b)** 2D FFT with two distinct peaks: The most prominent peak with flux periodicity $\Delta\boldsymbol{\Phi}\sim(\mathbf{3/2})\boldsymbol{\Phi}_\mathbf{0}$ corresponds to the interference of 'bunched' QPs with charge 2$e$/3. The weaker peak with flux periodicity $\Delta\boldsymbol{\Phi}\sim\mathbf{3}\boldsymbol{\Phi}_\mathbf{0}$ corresponds to the interference of elementary QPs with a charge $e$/3, suggesting partial dissociation of the bunched QPs. Note that the modulation gate voltage periodicity ($\Delta V_{MG}$) is almost twice as large in the dissociated condition. For a negative TG potential, see the main text.

## S9. Analysis of the deconvoluted peaks using the Gaussian filtering method

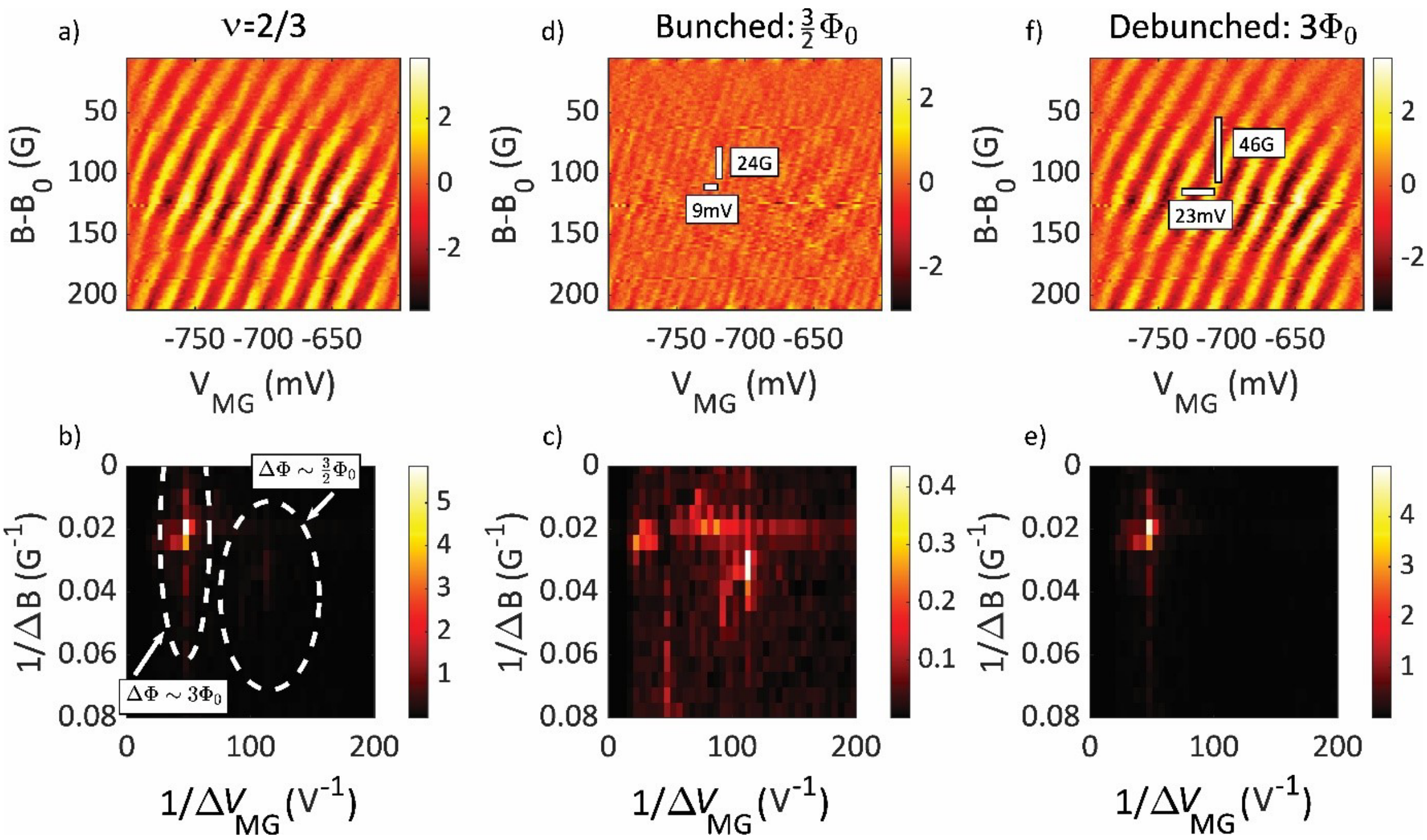

**Figure S9a. Separate views of flux periodicities at filling ν=2/3 with charged top gate. (a)** A strong peak at $\Delta\Phi=3\Phi_0$ (dissociated QPs) and a residual peak at $\Delta\Phi\cong(3/2)\Phi_0$ (bunched QPs). **(b)** Two corresponding peaks in the FFT. The dotted ovals indicate the radius of the Gaussian filtering applied later. **(c)** Removing the dissociated FFT peak at $3\Phi_0$ by Gaussian filtering and leaving the weaker peak of the bunched QPs. **(d)** Returning to *real space* pajama of the bunched QPs, with periodicity $\Delta\Phi\cong(3/2)\Phi_0$. **(e)** Filtering the residual bunched FFT peak. **(f)** Returning to *real space* pajama of the dissociated ('debunched') elementary QPs, $\Delta\Phi\cong3\Phi_0$.

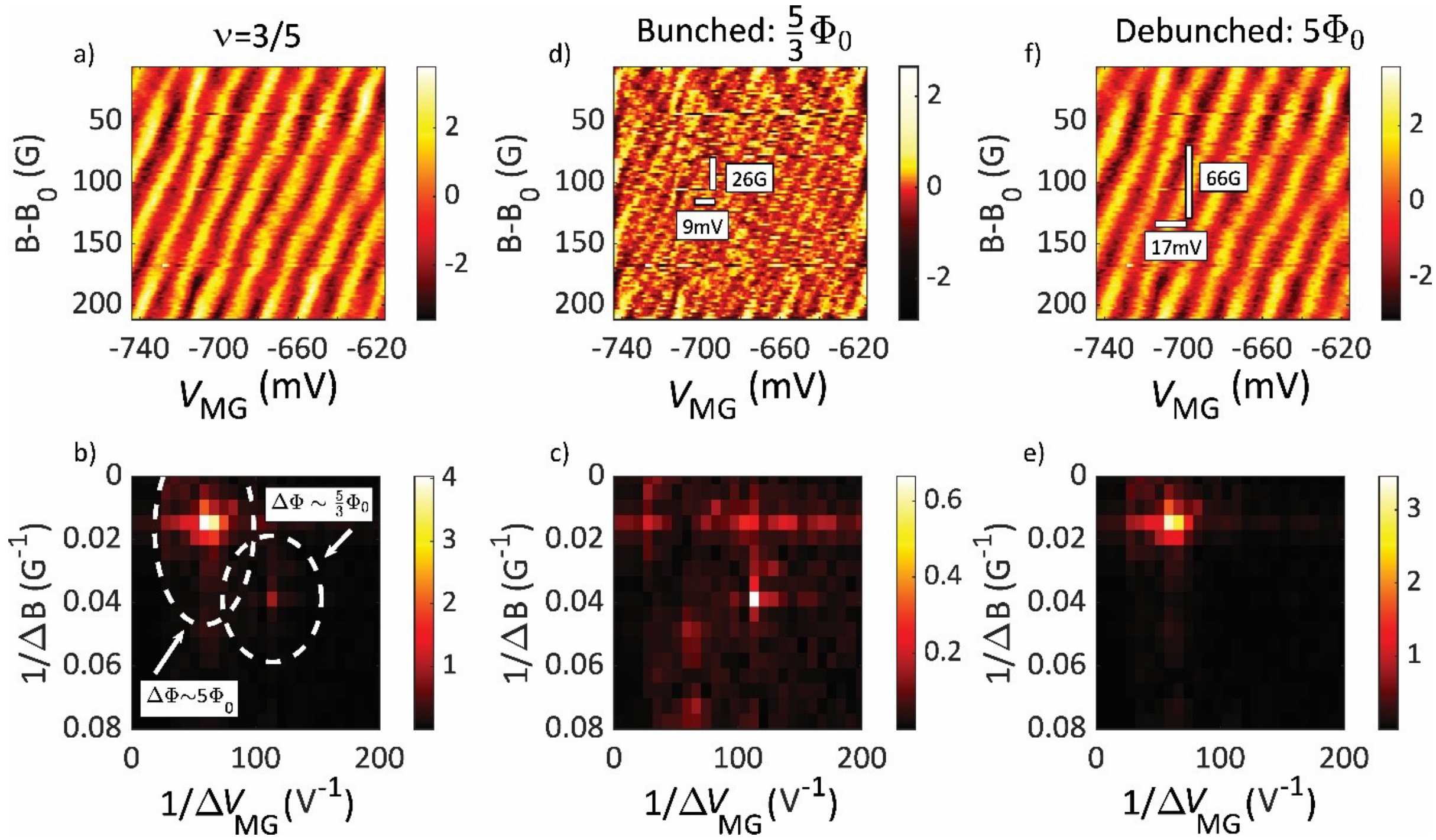

**Figure S9b. Separate views of flux periodicities at filling ν=3/5 with charged top gate. (a)** A strong peak at $\Delta\Phi \cong 5\Phi_0$ (dissociated QPs) and a residual peak at $\Delta\Phi \cong (5/3)\Phi_0$ (bunched QPs). **(b)** Two corresponding peaks in the FFT. The dotted ovals indicate the radius of the Gaussian filtering applied later. **(c)** Removing the dissociated FFT peak at $5\Phi_0$ by Gaussian filtering and leaving the weaker peak of the bunched QPs. **(d)** Returning to *real space* pajama of the bunched QPs, with periodicity $\Delta\Phi \cong (5/3)\Phi_0$. **(e)** Filtering the residual bunched FFT peak. **(f)** Returning to *real space* pajama of the dissociated elementary QPs, $\Delta\Phi \cong 5\Phi_0$.

**S10. The effect of temperature on the interference**

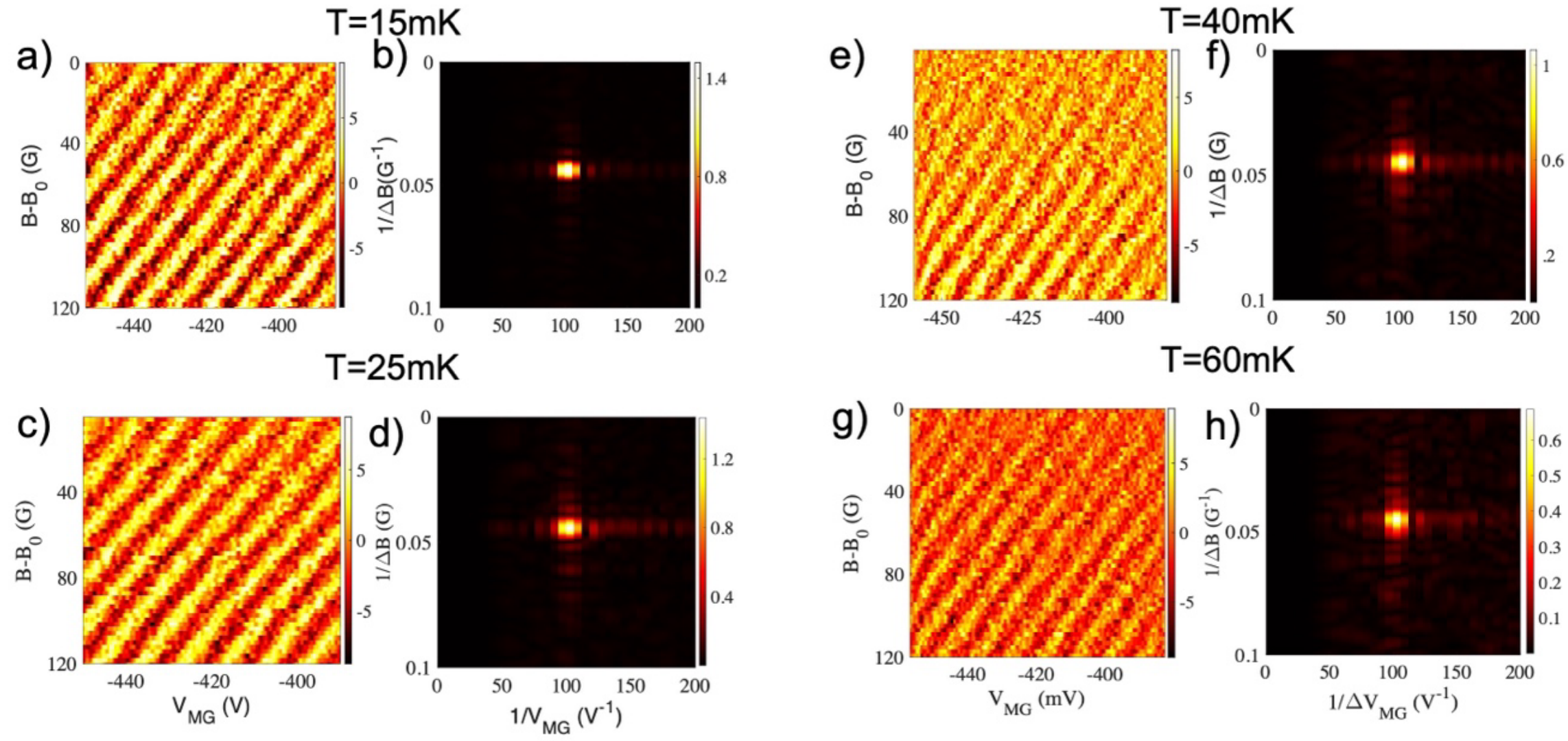

**Figure S10. The effect of temperature increase on the bunching of QPs tuned to ν$_b$=3/5.** AB interference at an elevated electron temperature in the range 15-100mK with QPCs transmission of: $t_1 = t_2$=0.8. In the range 15-60mK the visibility remains nearly constant (~1%) while the 2D-FFT plots remain with a single prominent peak of bunched QPs, $\Delta\Phi \sim (5/3)\Phi_0$. Above 60mK, the visibility drops abruptly. Similar measurements were also performed at ν=2/3 (not shown in the figure), showing qualitatively similar results. The decreasing visibility at relatively high temperatures prevents an observation of the dissociation.

**S11. Bunching and dissociation of the interference in the OMZI2 with a smaller top-gate (TG)**

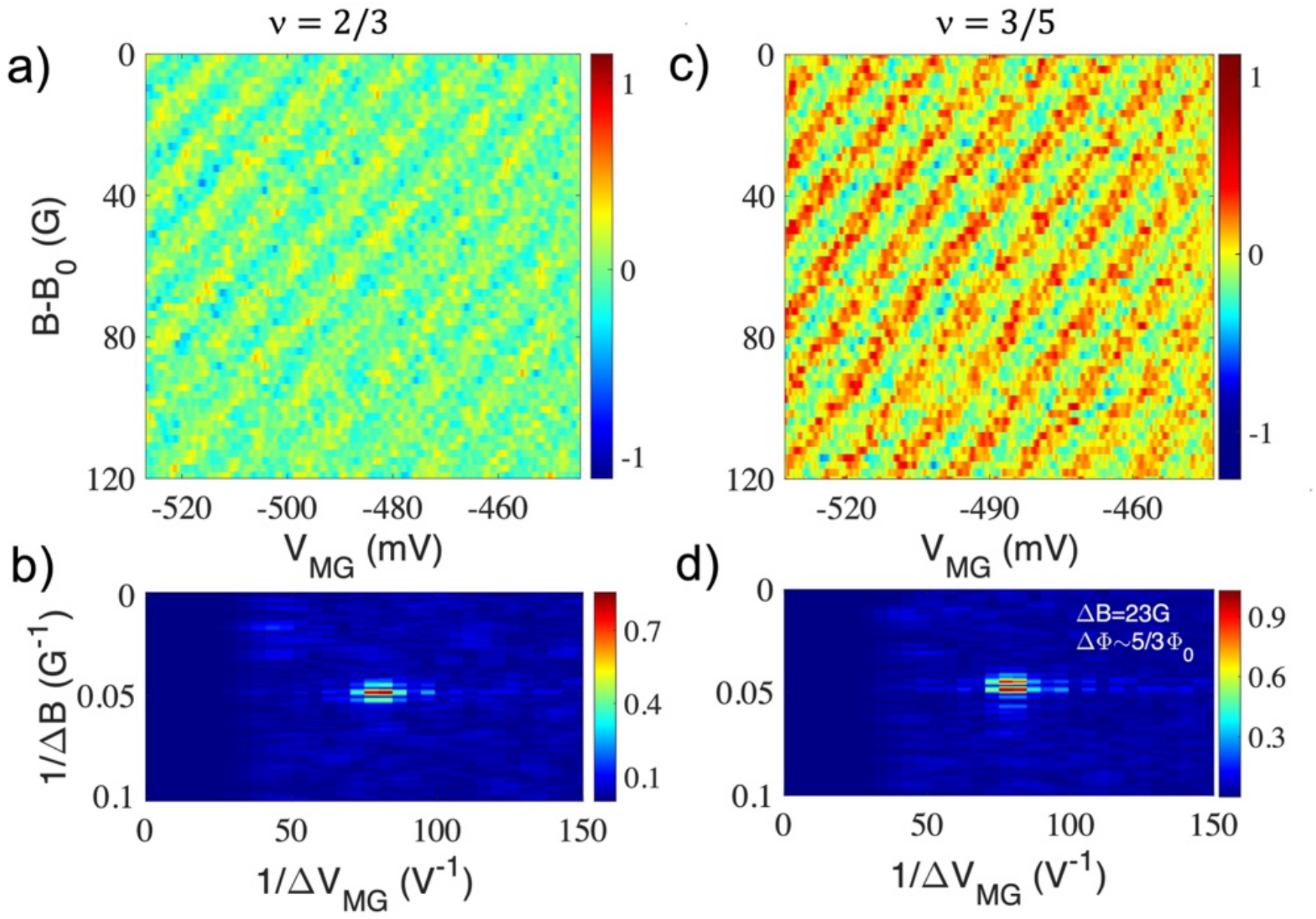

**Figure S11a. Aharonov Bohm interference at $\nu_b$=2/3 and 3/5 – Bunching.** The OMZI was fabricated on the same wafer. The interferometer area ($\sim 3\mu m^2$) and path length ($\sim 3\mu m$) are similar to the device described in the text. However, the transmission of the QPCs higher, and the top-gate area is half of the previous one. Here, $V_{TG}$=0V, with interference of bunched QPs at $\nu_b$=2/3 **(a, b)** and $\nu_b$=3/5 (**c, d**). The $B$-$V_{MG}$ interference is similar to the main device.

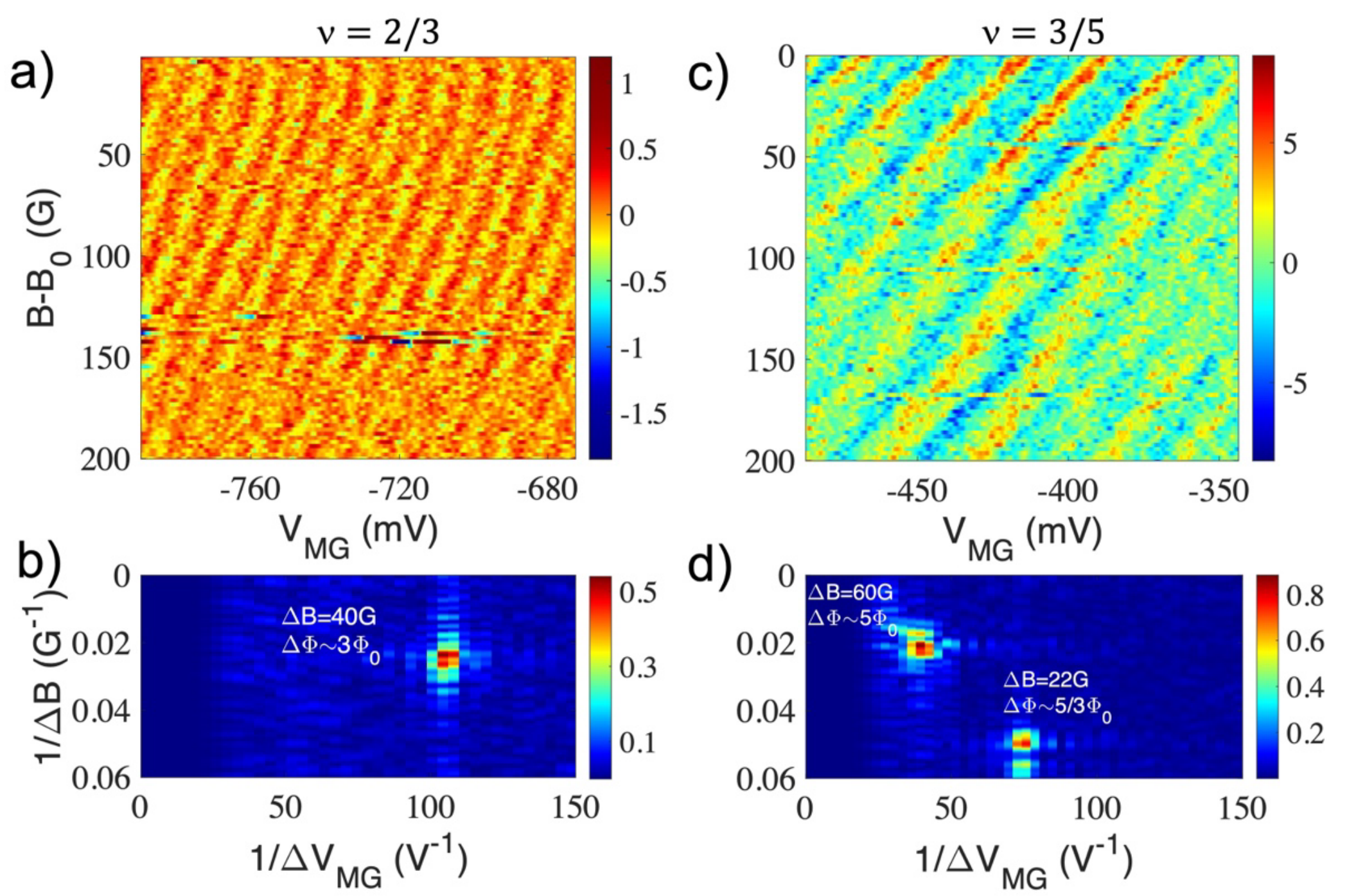

**Figure S11b. Aharonov Bohm interference at $\nu_b$=2/3 and 3/5 – Dissociation.** Aharonov Bohm oscillation at $\nu_b = 2/3$ and 3/5 for the filling factor configurations: **(a)** '1-2/3-0' **(c)** '1-3/5-0'. For this measurement, the top-gate was charged at $V_{TG}$=+(0.03-0.05)V and the QPCs were tuned to transmission of 90%. 2D FFT shows magnetic flux periodicity corresponding to $\Delta\Phi \cong 3\Phi_0$ for $\nu_b = 2/3$ in **(b)** and $\Delta\Phi \cong 5\Phi_0$ for $\nu_b = 3/5$ in **(d)** - suggesting a dissociation of the QPs. Additionally, a small residual harmonic, $\Delta\Phi \cong (5/3)\Phi_0$, is indicative of bunched QPs, is present at $\nu_b$=3/5. **(d)** However, this harmonic is absent for $\nu_b = 2/3$ **(b)** (likely due to its low visibility).

## S12. Quasiparticles and neutral modes in Abelian quantum Hall states

The boundaries of fractional quantum Hall states host one-dimensional edge modes that reflect the bulk topological order. All quantum Hall states host a chiral charge mode, dictated by the quantized Hall conductance in the bulk. In addition, there may be neutral modes propagating in the same direction ('downstream') or the opposite direction ('upstream'). Transferring a bulk quasiparticle to the boundary excites edge states in the form of quantized solitonic waves. Depending on the quasiparticle, the resulting edge excitation may or may not involve the neutral modes.

A well-known example is the $\nu = 2/3$ edge with vacuum **[1].** Without equilibration at the edge, it hosts a $\nu = 1$ mode $\phi_1$ and a counterpropagating $\nu = 1/3$ mode $\phi_2$. Charge equilibration due to disorder replaces the two modes with a total charge mode $\varphi_c = \phi_1 - \phi_2$ and a neutral mode $\varphi_n = \phi_1 - 3\phi_2$. A bulk quasiparticle with charge $e^* = e/3$ corresponds to a $-\frac{2\pi}{3}$ soliton in $\phi_2$, exciting both charge and neutral modes. An integer charge (an electron) can be encoded by a $2\pi$ soliton in either $\phi_1$ or $\phi_2$, which also excites both charge and neutral modes. However, combining a $-e^*$ quasiparticle ($\frac{2\pi}{3}$ soliton in $\phi_2$) and an electron ( $2\pi$ soliton in $\phi_1$) results in a $2e/3$ excitations for which the neutral mode is not excited, i.e., $2\pi - 3 * \left(\frac{2\pi}{3}\right) = 0$.

In general, the bulk quasiparticles and the edge theories of Abelian quantum Hall states are efficiently captured using the K-matrix formalism **[2]**. An integer-valued $n \times n$ matrix $K$ and an $n$-component integer vector $\boldsymbol{t}$ encodes a state at bulk filling factor $\nu = \boldsymbol{t}^T K^{-1} \boldsymbol{t}$. All bulk quasiparticles of the topological order $(K, \boldsymbol{t})$ are specified by integer vectors $\boldsymbol{l}$ with the electric charge given by,

$$Q(\boldsymbol{l}) = \boldsymbol{t}^T K^{-1} \boldsymbol{l}. \qquad (1)$$

The vacuum interface of this quantum Hall state hosts $n$ edge modes $\phi_i$ satisfying the commutation relation,

$$\left[\phi_i, \partial_x \phi_j\right] = 2\pi\, i\, K_{ij}^{-1} \qquad (2)$$

They can be organized into a charge mode $\varphi^c$ and $n - 1$ neutral modes $\varphi_a^n$ given by,

$\varphi^c = \boldsymbol{t} \cdot \boldsymbol{\phi}$,

$\varphi_a^n = \boldsymbol{b_a} \cdot \boldsymbol{\phi}$,

where $\boldsymbol{t}$ is the charge vector introduced above and $\boldsymbol{b_a}$ are $n - 1$ linearly independent vectors satisfying $Q(\boldsymbol{b_a}) = 0$. The charge mode decouples from the neutral modes at an equilibrated edge.

An edge excitation corresponding to a bulk quasiparticle specified by $\boldsymbol{l}$ is created by the operator,

$$\psi(\boldsymbol{l}) = \exp\,(i\, \boldsymbol{l} \cdot \boldsymbol{\phi}\,). \qquad (3)$$

Tunneling of this quasiparticle across a QPC is described by $\exp\left(i\, \boldsymbol{l} \cdot \boldsymbol{\phi}_{\text{top}} - i\, \boldsymbol{l} \cdot \boldsymbol{\phi}_{\text{bottom}}\right)$, where $\boldsymbol{\phi}_{\text{top}}$ and $\boldsymbol{\phi}_{\text{bottom}}$ are the edge modes on opposite sides of the QPC. For a generic quasiparticle described by $\boldsymbol{l}$, this operator exhibits non-trivial commutation relations with the charge modes **and** with the neutral modes according to Eq. (2). Specifically, it creates the solitonic waves discussed above. However, the Laughlin quasiparticle, which carries charge $e_\Phi = \nu e$ is described by $\boldsymbol{l}_{\text{Laughlin}} = \boldsymbol{t}$ commutes with *all* neutral modes due to $Q(\boldsymbol{b_a}) = 0$. **Consequently, tunneling of a Laughlin quasiparticle excites only the total charge mode, like the 2e/3 excitation in the example.**